\documentclass[11pt]{elsarticle}

\pdfoutput=1

\makeatletter
\def\ps@pprintTitle{%
  \let\@oddhead\@empty
  \let\@evenhead\@empty
  \let\@oddfoot\@empty
  \let\@evenfoot\@oddfoot
}
\makeatother

\usepackage{url}
\usepackage{breakurl}
\usepackage[breaklinks,
            colorlinks = true,
            linkcolor = blue,
            urlcolor  = blue,
            citecolor = blue,
            anchorcolor = blue]{hyperref}

\usepackage{lineno}

\usepackage{graphicx}
\usepackage[margin=1.25in]{geometry}
\usepackage[usenames,dvipsnames]{color}
\newcommand\acp{\begin{center}
\rule[-0.2in]{\hsize}{0.01in}\\\rule{\hsize}{0.01in}\\
\vskip 0.1in Submitted to the  Proceedings\\ 
of the African Conference on Fundamental and Applied Physics
    \vskip 0.05in
    {\it Second Edition, ACP2021, March 7--11, 2022 --- Virtual Event}\\
\rule{\hsize}{0.01in}\\\rule[+0.2in]{\hsize}{0.01in} \\
\end{center}}

\usepackage[firstpage=true]{background}
\backgroundsetup{contents={\parbox{6.5in}{\acp}}, scale=1,placement=top,opacity=1,color=black,position={3.25in,1.2in}}

\usepackage{fancyhdr}
\fancypagestyle{plain}{%
  \fancyhf{}%
  \fancyhead[C]{}
  \fancyfoot[C]{\thepage}
}

\fancypagestyle{empty}{%
  \fancyhf{}%
  \fancyhead[C]{{\it ACP2021, March 7--11, 2022 --- Virtual Edition}}
  \fancyfoot[C]{\thepage}
}
\emergencystretch=\maxdimen
\begin{document}

\begin{frontmatter}

%\pubblock

\title{Young Physicists Forum and the Importance for Education and Capacity Development for Africa}

\author[add1]{Benard Mulilo\corref{cor1}}
\ead{benard.mulilo@unza.zm}
\author[add2]{Mounia Laassiri}
\author[add3]{Diallo Boye}

\cortext[cor1]{Corresponding Author}

\address[add1]{University of Zambia, Zambia}
\address[add2]{Mohammed V University, Morocco}
\address[add3]{Brookhaven National Laboratory, US}

\begin{abstract}
\noindent 
Higher education and advanced scientific research lead to social, economic, and political development of any country. All developed societies like the current 2022 G7 countries: Canada, France, Germany, Italy, Japan, the UK, and the US have all not only heavily invested in higher education but also in advanced scientific research in their respective countries. Similarly, for African countries to develop socially, economically, and politically, they must follow suit by massively investing in higher education and local scientific research.  
\end{abstract}


\end{frontmatter}

%%%%%%%%%%%%%%
%\linenumbers
%%%%%%%%%%%%%%

%\def\thefootnote{\fnsymbol{footnote}}
%\setcounter{footnote}{0}

%\newpage

\section{Introduction}
\label{sec:intro}
\noindent
In 2009, the United Nations Population Fund announced that the population of Africa had reached the one-billion mark and doubled in size in 27 years~\cite{undp}. Regardless of the size and large pool of the human resource that the continent is endowed with, most African countries still continue struggling economically. Based on World Bank estimates~\cite{worldbank}, the proportion of Africans living on less than US\$ 1.90 per day fell from 56\% in 1990 to 43\% in 2012. This indicates a positive improvement of 13\% in the living standards of people in Africa though according to the World Bank Report~\cite{worldbank}, there were still more poor people in Africa in 2012 than in 1990 estimated to be more than 330 million up from about 280 million due to rapid population growth~\cite{undp} that the continent has been undergoing over the years. Furthermore, despite poverty being a major problem in Africa~\cite{worldbank}, the continent also experiences deadly diseases such as the Acquired Immune Deficiency Syndrome (AIDS) caused by the Human Immunodeficiency Virus (HIV) believed to have originated from Africa~\cite{hivorigin2,hivorigin}; Ebola virus disease~\cite{ebola} whose fatality rate is around 50\% with case fatality rates ranging from 25\% to 90\% in past outbreaks~\cite{ebola}, and the recent outbreak of the COVID-19 pandemic~\cite{covid}, which has impacted negatively on Africa and the rest of the world. The continent also faces challenges of science and technology~\cite{technology} with many African countries technologically depending on other continents for engineering, educational, agricultural, and health services, among others. African countries also face inadequate research-output capability or interest with Africa noted to generate only less than 1\% of the world's research output~\cite{research} despite its increasing population~\cite{undp}. Due to all these challenges and other factors, the continent has seen young, talented, skilled, and educated Africans leaving African countries in search for better opportunities overseas, a trend referred to as brain drain~\cite{braindrain}. To overcome all these challenges and others, African countries should emulate the scientific, education, health, social, political, and economic policies of developed societies such as the Group of Seven (G7) 2022 countries: Canada, France, Germany, Italy, Japan, the UK, and the US~\cite{g7}, which have all massively invested in higher education, science, and technological advancements. The Young Physicists Forum (YPF)~\cite{ypf} was founded in 2021 by the African Strategy for Fundamental and Applied Physics (ASFAP)~\cite{asfap}, amid the COVID-19 pandemic~\cite{covid}, to identify the major challenges that young physicists face and solutions thereof in order to positively contribute to the educational and local-scientific research on the continent, and thus, build capacity for Africa. 

\section{Young Physicists Forum}
\label{sec:grate}
\noindent

\begin{figure}[htbp]
	\centering
	\includegraphics[width=1.00\linewidth]{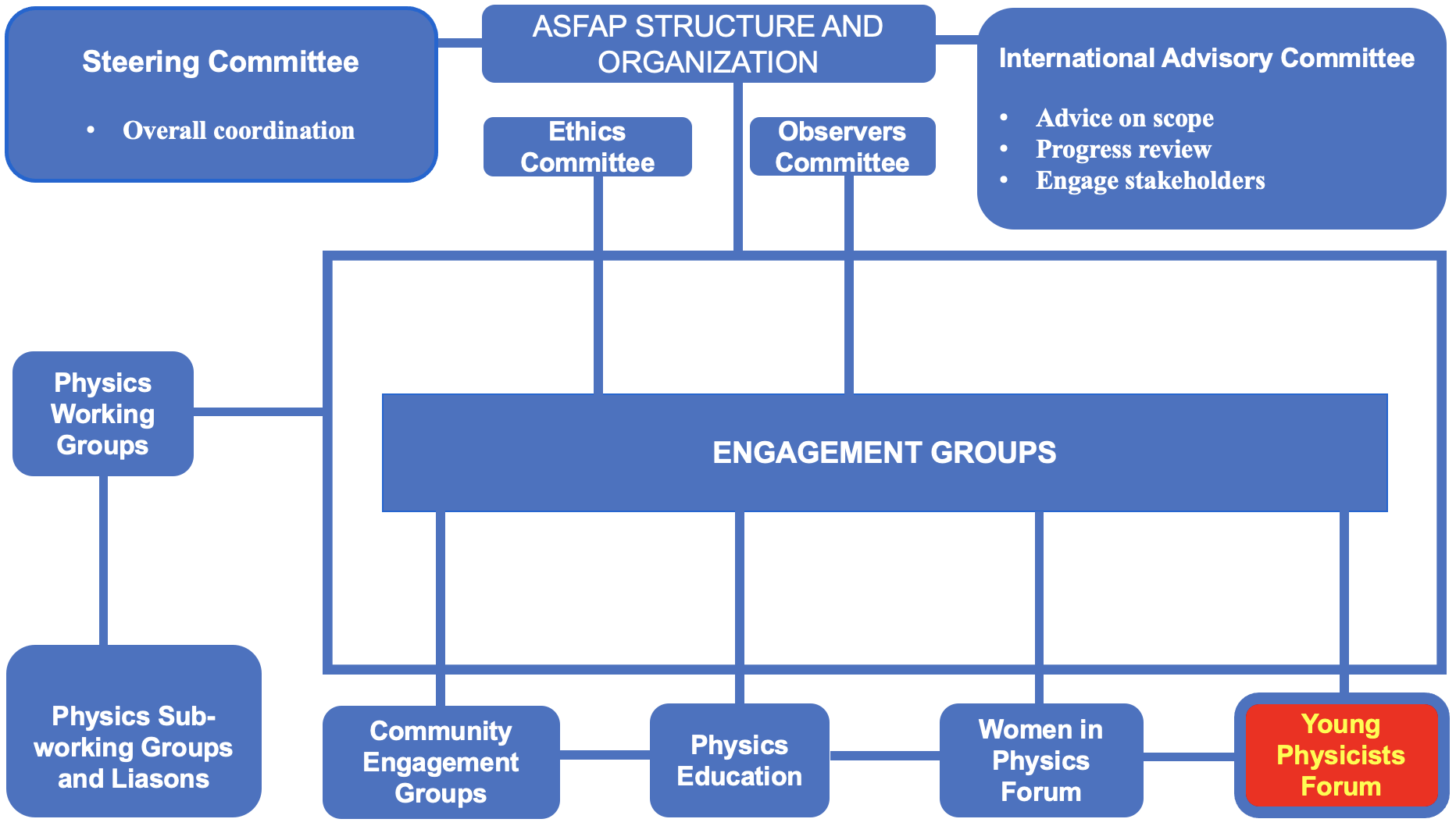}
	\caption{Structure and organization of the African Strategy for Fundamental and Applied Physics.}
	\label{asfap}
\end{figure}

The Young Physicists Forum~\cite{ypf} is one of the engagement and physics working groups (PWG) under the African Strategy for Fundamental and Applied Physics (ASFAP)~\cite{asfap}. The forum is driven by three, young, and vibrant physicists who are co-conveners of the group all in possession of a doctor of philosophy in physics \cite{ypf}. The co-conveners' mandate is, among other things, to ensure that the group remains sharply focused on its aims and objectives. The forum has a total of 65 active and registered members~\cite{ypf}, most of whom are in possession of either a master of science degree or doctor of philosophy in physics. There is, however, no discrimination regarding the highest level of education YPF members~\cite{ypf} must meet and, therefore, all interested individuals within and outside the African continent are eligible to join the forum~\cite{ypf} as long as they sign up~\cite{ypf} and get approved by the steering committee of ASFAP~\cite{asfap}. The group also encourages bachelors students in various science disciplines, particularly physics, from various African universities to join the YPF~\cite{ypf} and enjoy the mentoring/scholarship benefits that YPF members share within the group, and thus increase their chance of embarking on postgraduate studies either within Africa or overseas. The Young Physicists Forum~\cite{ypf} reports to the steering committee of ASFAP~\cite{asfap} in a well organized structure as shown in Figure~\ref{asfap}. 

The aims and objectives of the ASFAP-YPF~\cite{ypf} are, among others, to collect ideas, opinions, and experiences on education, physics outlook, careers, workplace environment, and scientific research in Africa. Furthermore, the forum is mandated to clearly identify and raise awareness of the educational challenges and science career opportunities for young physicists in Africa and advocate for change by informing policymakers for action. Last, but not the least, the forum also aims at collecting preliminary data for future research. Since the group's inception in 2021, the Young Physicists Forum~\cite{ypf} has made tremendous progress in meeting its mandate (i.e., its aims and objectives) with the main modes of information dissemination being through scheduled meetings within the group and regular co-conveners' meetings, which are usually held on a weekly basis on Wednesday at 5:00 PM, Coordinated Universal Time (UTC). The forum has also formulated a survey~\cite{survey} to solicit for a wider community input of ideas. In addition, the group has so far held one successful workshop with stakeholders within and outside ASFAP~\cite{asfap} that was virtually conducted on 26$^{th}$ January, 2022 tagged \textit{ASFAP: YPF-Challenges and Opportunities}~\cite{workshop}. The YPF~\cite{ypf} also actively participated in the second edition of the African Conference on Fundamental and Applied Physics tagged \textit{ACP2021}~\cite{ACP2021-report} and contributed three talks under different themes mainly focused on the status and progress the forum has so far made in line with the aims and objectives of the group.

\section{Challenges and Opportunities Survey}
\begin{figure}[htbp]
	\centering
	\includegraphics[width=1.00\linewidth]{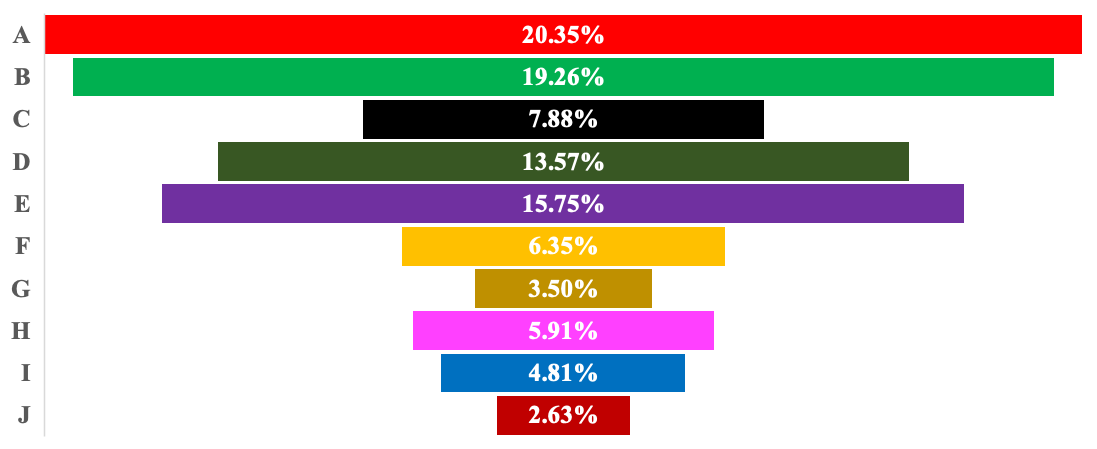}
	\caption{Challenges faced by respondents pursuing their highest level of education in African universities.}
	\label{challenges}
\end{figure}

To solicit for a wider community input, the Young Physicists Forum~\cite{ypf} has opened a survey~\cite{survey} to sample African respondents within and overseas, main of whom are alumni of the African School of Physics (ASP)~\cite{ASP}. The survey~\cite{survey} is aimed at gathering information on the education background, research performance, collaboration opportunities, career development, and workplace environment of the respondents. Survey results~\cite{survey} show that 79.56\%, of the respondents pursued their highest level of education within Africa while 20.44\% of the respondents attained their highest level of education outside the continent of Africa. The survey~\cite{survey} has further revealed that of the respondents who attained their highest level of education within Africa normalized to 100\%, only 39.42\% were satisfied. Factors leading to the educational dissatisfaction rate by respondents are plotted in Figure~\ref{challenges} and outlined in Table~\ref{ed-challenges}. From Figure~\ref{challenges} and Table~\ref{ed-challenges}, it is evident that good quality education and research in Africa still remain a huge challenge. Other major obstacles of an African educational system include the lack of mentors, skills training, libraries, job insecurity, and to a lesser extent political instability such as wars, among others. Since education, science, and technology are ingredients that contribute massively to good life and development of the global economy, there is need to solicit for remedies that counter the education and research challenges that many African countries have been grappling with for years. 

\begin{table}[htbp]
\centering
\caption{Educational challenges faced by respondents pursuing higher education in African institutions}
\hspace{0.5cm}
\begin{tabular}{|p{3cm}|p{8.0cm}|p{3.0cm}|} \hline\hline
Responses & Challenges & Rate (\%) \\ \hline\hline 
A & Lack of research funding & 20.35\\ \hline
B & Lack of research equipment & 19.26\\ \hline
C & Lack of mentoring support & 7.88\\ \hline
D & Lack of mobility opportunities & 13.57 \\ \hline
E & Lack of proper skills training & 15.75\\ \hline
F & Lack of access to libraries & 6.35\\ \hline
G & Limitation of academic freedom & 3.50\\ \hline
H & Imbalance between work and family demands & 5.91\\ \hline
I & Job insecurity & 4.81\\ \hline
J & Political instability and wars & 2.63\\ \hline
\end{tabular}
\label{ed-challenges}
\end{table}

According to the survey~\cite{survey} being conducted by the Young Physicists Forum~\cite{ypf}, prominent solutions to educational challenges include raising awareness to African policymakers and private enterprises on the need to fund research through provision of grants, which universities in Africa should accountably utilize to buy experimental equipment and conduct meaningful research. African governments should also invest in building higher learning institutions that are well equipped with research facilities such as modern laboratories where academic staff and their students could establish the link between theory and experimental work. This would then help reduce over-dependence on foreign research facilities and contribute to meaningful and solid collaboration relationships with other institutions and research facilities overseas. Public and private universities should work together and help improve the internet network in universities and research facilities across Africa as a good and stable internet connectivity undoubtedly enhances scientific research output and helps improve the quality of learning.

\begin{figure}[htbp]
	\centering
	\includegraphics[width=1.00\linewidth]{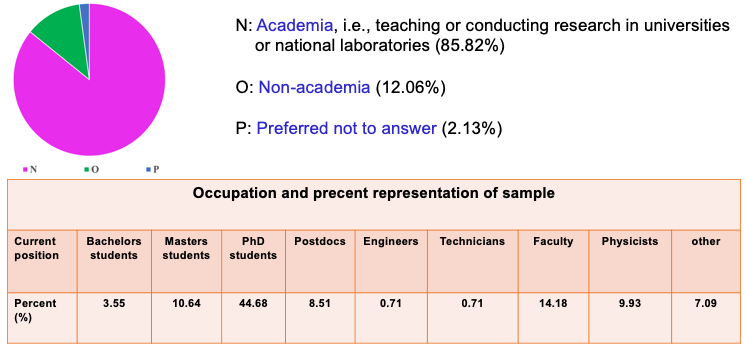}
	\caption{Occupation and percent representation of respondents according to the survey conducted by YPF.}
	\label{educational}
\end{figure}

Other measures that may help counter educational challenges in Africa include revision of the school and university curricula by reducing over-dependence on theoretical work~\cite{survey}, building scientific research facilities, and securing laboratory equipment to encourage research skills and knowledge acquisition through experimental work among African students. Furthermore, the lack of mentors in science disciplines like physics in African universities could be resolved by motivating professors to embark on scientific-research projects and closely working with their students~\cite{survey} once research grants are available to them from governments and private enterprises. Academic staff should also spend more advisory time with their students and try and establish the link between theoretical and experimental work together~\cite{survey}. Additionally, academic staff members should offer more structured feedback to students and also establish research collaborations within and outside the continent so as to expose their students scientifically~\cite{survey}. Occupational and career guidance should also be provided to students by their advisors in order to motivate them regarding their future endeavours in academia within Africa~\cite{survey}. Career with occupational development is another huge challenge being faced by young physicists in Africa~\cite{survey}. According to the population sampled in the survey~\cite{survey}, it is found that roughly 85.82\% of the respondents are in the field of academia where they are teaching and conducting research in national universities and laboratories while those in non-academia fields accounted to about 12.06\%, and approximately 2.13\% preferred not to reveal there occupation as shown in the pie chart in Figure~\ref{educational} by N, O, and P, respectively. Those in academia identified themselves as bachelors, masters, and doctoral students, postdocs, engineers, technicians, physicists as well as faculty members.

\begin{figure}[htbp]
	\centering
	\includegraphics[width=1.00\linewidth]{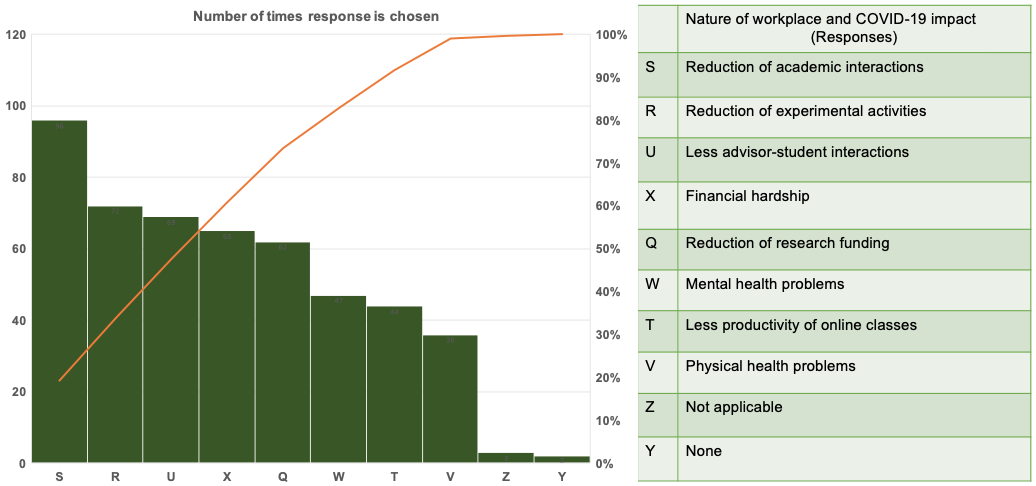}
	\caption{Impact of the nature of the workplace and COVID-19 pandemic on research institutions in Africa.}
	\label{workplace}
\end{figure}

Results of the survey~\cite{survey} have further revealed that securing an academia position in universities and national research facilities within Africa poses a major challenge and is, at the same time, a huge sacrifice owing to the fact that the workplace environment is mostly not conducive due to lack of experimental equipment, among other challenges, more so in the last two years with the breakout of the COVID-19 pandemic~\cite{covid}. Based on the results of the survey~\cite{survey}, the Young Physicists Forum~\cite{ypf} have learnt that the effects of the nature of an academia-workplace environment in Africa and the impact of the COVID-19~\cite{covid} have led to a reduction of academic interactions between academic staff and students, experimental activities, and research funding as highlighted in Figure~\ref{workplace}. Other negative effects include less advisor-student interactions, physical and mental-health problems as well as financial hardships, among other challenges, as described in Figure~\ref{workplace}. The poor currency-exchange rate of African currencies against major world currencies such as the United States Dollars ($\$$) and British Pound ($\pounds$), among others, is another major challenge~\cite{survey} not explicitly stated in Figure~\ref{workplace}, but being faced in the academia field in Africa as this significantly and negatively impacts scientific-collaboration work between Africa and other continents as far as international research visits and conferences by students and academic staff are concerned.

The lack of good will and minimal interest in education, science, and technology in Africa~\cite{technology} have led to a huge challenge over the years where the world has witnessed a large number of skilled manpower leaving Africa for other continents in search of a more conducive workplace environment and an attractive income to support their families, a trend known as brain drain~\cite{braindrain}. The survey~\cite{survey} being conducted by the YPF~\cite{ypf} has so far revealed some instances of brain drain~\cite{survey,braindrain} that have been taking place in Africa over the years. These include young and skilled African students studying abroad on scholarships opting to stay and work overseas after completion of their studies~\cite{survey}. Researchers and postdocs also feel more comfortable working overseas than in African universities where they are either not welcomed or because of the nature of an African academia workplace environment and meagre salaries~\cite{survey}. The lack of academic freedom (i.e., students having no choice of what to study due to financial reasons), inadequate funding, and absence of research equipment disfavor Africa as a good destination for good quality education and research work~\cite{survey}. Political instability such as wars in some countries in Africa drive away academically qualified personnel to other countries outside the continent where they settle down peacefully and continue to contribute to science and technology there than in their African countries of origin~\cite{survey}. In spite of all these brain drain challenges~\cite{survey,braindrain}, the hope for Africa in education, science, and technology~\cite{technology} is still alive. Through the survey~\cite{survey}, the YPF~\cite{ypf} have come up with measures to counter the effects of brain drain~\cite{braindrain} and hence help keep alive the hope of African countries to develop their education and build capacity for Africa. Some of these interventions are summarized and listed in Table~\ref{braindrain}. 

\begin{table}[htbp]
\centering
\caption{Measures that may help counter brain drain according to the survey conducted by the YPF \label{braindrain}}
\hspace{0.5cm}
\begin{tabular}{| p{1cm} | p{13.0cm}|}\hline\hline
 & Interventions that may help counter brain drain in most African countries \\ \hline\hline 
1 & Create a school of excellence within Africa for Africans who have obtained their baccalaureate with honors in order to encourage African academic excellence and experience.\\ \hline
2 & Policymakers on the continent should partner with private enterprises and work together to improve the research-workplace environment and conditions of service such as salaries to match foreign-based counterparts in academia.\\ \hline
3 & Create national research laboratories and more academic positions in African universities and provide research grants to enable academic staff members to embark on a meaningful scientific research experience within the continent. \\ \hline
4 &Policymakers should stabilize African currencies to compete favorably with other major world currencies such that the salaries skilled academic staff are earning in Africa are favourably comparable to salaries fellow counterparts earn abroad.  \\ \hline
5 & Enhance and connect African academic infrastructures with the rest of the world; promote scientific collaborations with international universities, research institutions, and laboratories and allow creative young Africans to present new scientific research projects.\\ \hline
6 & Massive investment in African university education is required that will result in an increase in well paying jobs. A marketing campaign should be setup to encourage the youth to stay and work in their respective countries in Africa.\\ \hline

\end{tabular}
\label{braindrain}
\end{table}

\section{Conclusions}
\label{2}

The African continent is endowed with abundant natural resources ranging from huge arable land through oil, natural gas, and minerals to floras and faunas. It is amazingly puzzling to note that the continent holds a large proportion of the world's natural resources, both renewable and non-renewable and yet, to a large extent, Africa still remains undeveloped with higher poverty levels~\cite{worldbank} than other continents. To restrain or minimize these challenges, Africa should heavily invest in higher education and promote local scientific research~\cite{survey,technology}. Advanced scientific research carried out within Africa would, for example, help find solutions to diseases such as HIV/AIDS~\cite{hivorigin,hivorigin2} that have been ravaging the continent over the years; produce vaccines of its own to cure pandemics such as COVID-19~\cite{covid} without having to entirely depend or solely wait for developed societies~\cite{g7} to share portions of their vaccines; process its abundant natural resources from raw materials to finished products, and reduce over-dependence on developed countries for finished goods and services~\cite{technology}. This would, in turn, build an even better relationship between Africa and the rest of the world as far as business is concerned. Since higher education and research are key to social, political, and economic independence of any country, it goes without saying that, education and research should be prioritized across Africa and ensure that educated human resource is enticed to stay and work within the continent by offering an attractive workplace environment and competitive conditions of service and thus, help minimize the brain-drain~\cite{survey,braindrain} phenomenon. The YPF~\cite{ypf} is entirely open and solely devoted to identifying the challenges that young physicists face in developing their careers in Africa and finding solutions as well as career opportunities available for young physicists on the continent so as to revamp education and build capacity for Africa. The YPF is also entirely committed to mentor young physicists in Africa and to help promote research collaborations with other young physicists globally~\cite{survey}. All in all, the YPF~\cite{ypf} is willing to partner with policymakers across the continent and beyond, the private sector, and business enterprises as far as promotion of higher education and advanced, local scientific-research projects in Africa are concerned.

\section*{Acknowledgments}
Young Physicists Forum co-conveners are thankful to all YPF members, ASFAP's Engagement and Physics Working Groups for their positive contributions. Gratitude also goes to ASFAP's steering committee and stakeholders who have made this work possible. 

%\newpage
%\section*{Appendix}

\newpage

\bibliographystyle{elsarticle-num}
\bibliography{myreferences} 

\end{document}